\documentclass[11pt]{article}

\usepackage[margin=1in]{geometry}
\usepackage{graphicx}
\usepackage{amsmath,amssymb}
\usepackage{booktabs}
\usepackage{hyperref}
\usepackage[numbers,sort&compress]{natbib}
\usepackage{subcaption}
\begin{document}

% ==============================
% Title
% ==============================
\begin{center}
{\LARGE \textbf{Optical superradiance from single-digit-femtosecond electron beam structure}}\\[0.75em]

{\small
Chad Pennington$^{1}$, Gia Azcoitia$^{1}$, Blae Stacey$^{2,3}$, Willi Kuropka$^{3}$,\\
Jackson Rozells$^{1}$, Francois Lemery$^{3}$, Florian Burkart$^{3}$, Sergio Carbajo$^{1,4,5}$
}\\[0.5em]

{\footnotesize
$^{1}$Department of Electrical and Computer Engineering, University of California, Los Angeles, CA 90095, USA\\
$^{2}$Universit\"at Hamburg, Department of Physics, 22607 Hamburg, Germany\\
$^{3}$Deutsches Elektronen-Synchrotron DESY, 22607 Hamburg, Germany\\
$^{4}$Department of Physics and Astronomy, University of California, Los Angeles, CA 90095, USA\\
$^{5}$SLAC National Accelerator Laboratory, Menlo Park, CA 94025, USA
}
\end{center}

\vspace{1em}

% ==============================
% Abstract
% ==============================
\section*{Abstract}

We report measurements of superradiant optical transition radiation in the 550-800 nm range produced by ultrashort relativistic electron bunches at a dielectric boundary. In the measured optical spectra, we observe photon production with quadratic charge dependence in the visible range, consistent with optical frequency coherence determined by the longitudinal electron bunch form factor. The measured spectral envelope is reproduced by a theoretical model of coherent transition radiation (CTR), which is consistent with a sub-femtosecond longitudinal feature within the electron bunch with characteristic scale $\tau_{\mathrm{FWHM}} = 1.2~\mathrm{fs}$. These results extend CTR from the terahertz into the visible spectrum without the use of undulators or externally seeded microbunching. This superradiant boundary emission in the optical range opens a route to tunable coherent radiation from charged particle beams and provides a platform for broadband coherent light generation, enabling new opportunities for phase-sensitive optical experiments.% ==============================
\section{Introduction}
% ==============================

Transition radiation occurs when a charged particle crosses an interface between materials with different dielectric properties, producing broadband electromagnetic emission that reflects the particle's instantaneous electromagnetic field \cite{GinzburgFrank1945,TerMikaelian1972}. For relativistic electron beams this radiation, known as optical transition radiation (OTR), is widely used as a diagnostic tool because it enables single-shot measurements of beam position, size, and divergence with optical resolution \cite{FioritoRule1994}. In the incoherent regime the emitted intensity scales linearly with the number of electrons in the bunch.

When the longitudinal bunch length becomes comparable to or shorter than the emitted wavelength, the radiation fields from individual electrons add coherently and the emitted intensity scales quadratically with bunch charge. This regime, known as Coherent Transition Radiation (CTR), provides a sensitive probe of ultrafast beam structure through the longitudinal bunch form factor \cite{GoldsmithWalsh1988,Lihn1996,Shibata1999}. Because the coherence condition is readily satisfied for sub-picosecond beams at terahertz and mid-infrared wavelengths, CTR has primarily been exploited in these spectral regions for bunch length diagnostics in free-electron lasers, laser–plasma accelerators, and ultrafast electron sources \cite{CarrEtAl2002,CasalbuoniEtAl2009,LumpkinEtAl2006}.

Extending coherent transition radiation into the optical regime requires electron bunches with longitudinal structure on femtosecond or sub-femtosecond timescales. Achieving and preserving such ultrashort structures in accelerator beamlines is challenging due to strong space-charge forces, collective effects, and sensitivity to longitudinal phase-space evolution. As a result, coherent optical transition radiation (COTR) has most commonly been observed in systems where density modulations arise from microbunching instabilities or free-electron laser processes \cite{Lumpkin2009PRSTAB, Lumpkin2002PRL, Lumpkin2020PRL, HuangEtAl2004}.

Despite these challenges, COTR occupies a uniquely valuable regime at the intersection of coherent radiation physics
and optical diagnostics. By bridging conventional incoherent OTR and long-wavelength CTR, COTR enables direct
access to ultrafast beam structure with optical spatial resolution and standard imaging and spectroscopic tools
\cite{Ferrario2010,Xiang2012,Musumeci2010}. Unlike terahertz CTR diagnostics, which often require specialized
detectors and optics, optical-frequency radiation can be readily collected, dispersed, and imaged using
well-established instrumentation. Moreover, the presence of temporal coherence at optical wavelengths opens new
opportunities for probing femtosecond-scale density modulations, longitudinal microstructure, and collective beam
dynamics that are otherwise difficult to resolve.

Beyond diagnostics, COTR represents a novel mechanism for generating broadband coherent optical radiation from
accelerator systems. In contrast to undulators or inverse Compton sources, COTR relies solely on boundary
interactions and ultrashort bunch formation, potentially enabling highly tunable light sources in the visible and
near-infrared \cite{Rosenzweig2018,Papadopoulos2020,Tantawi2020}. Recent advances in high-gradient photoinjectors and ultralow-emittance beam generation have begun to push electron sources into regimes
where optical coherence becomes experimentally accessible \cite{BrownellWalshDoucas1998,GoverEtAl2019}, motivating interest in COTR as a diagnostic and radiation source.

Here we report the observation of superradiant optical transition radiation produced by single-digit femtosecond electron bunches at a dielectric boundary. By measuring wavelength-resolved charge scaling of the emitted optical spectra, we observe quadratic charge dependence in the visible range. The observed dependence implies longitudinal beam structure on femtosecond or sub-femtosecond timescales. Strong longitudinal compression in the accelerator beamline is expected to produce fine-scale density structure within the bunch that can satisfy the optical coherence condition. A coherent transition radiation model reproduces the measured spectral envelope and yields a characteristic bunch duration of $\tau_{\mathrm{FWHM}} \approx 1.2$ fs.
These results demonstrate optical-frequency superradiance from compressed electron bunches without a free-electron laser (FEL) or externally seeded density modulation, extending coherent transition radiation into the visible regime and establishing boundary-driven superradiance as a new mechanism for generating broadband coherent optical radiation from relativistic electron beams. While fine-scale longitudinal structure may arise during compression, no externally seeded or FEL-driven microbunching mechanism is present in the beamline.

% ==========================================================
\section{Methods}
% ==========================================================

\subsection{Electron Beam Source}
The experiments were carried out at the ARES (Accelerator Research Experiment at SINBAD) facility at Deutsches Elektronen-Synchrotron (DESY) in Hamburg \cite{BurkartEtAl2022}. ARES is a dedicated research linear accelerator designed to produce ultrashort, high-brightness electron bunches for advanced accelerator and beam physics studies. 

A state of the art S-band photoinjector produces electron bunches with up to 50 Hz repetition rate. Longitudinal compression is achieved using a movable magnetic bunch compressor \cite{KuropkaEtAl2022} located downstream of two traveling wave structures (TWS1 and TWS2). Using 120 MeV kinetic energy, energy chirp, and longitudinal dispersion optimized for compression, an ultra-short bunch length was confirmed using an X-band transverse deflecting cavity \cite{Grudiev2016} at the same longitudinal position in the machine as the CTR target. The bunch length measurement was resolution limited to ca. 4 fs rms.

In the experiment the beam was focused onto a scintillating screen, which is installed on the same mover system as the CTR target. The low emittance of the beam allows transverse dimensions of a few micrometers to be achieved, although the beam size measurement was limited by the screen’s resolution of ca. 10 um. In the wavelength regime of interest also the transverse form factors become important parameters of the radiation properties \cite{KubeJINST2018}.

\begin{figure}[ht]
\centering
\includegraphics[width=\textwidth]{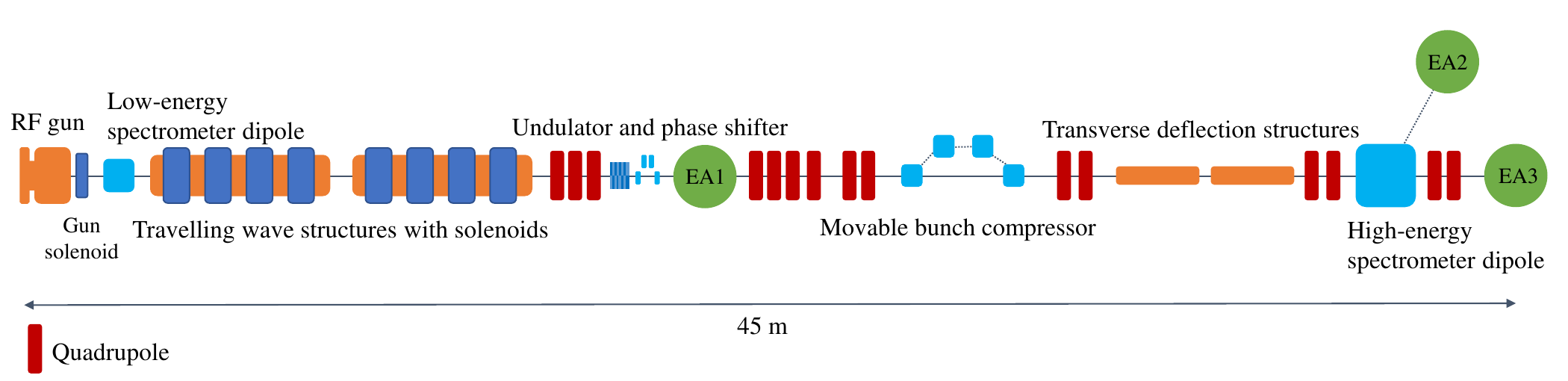}
\caption{Schematic of the ARES linear electron accelerator. Taken from \cite{JasterMerz2023Thesis}.}
\label{fig:aresschematic}
\end{figure}

CTR was generated when the relativistic electron beam impinged on a silver mirror mounted at an incidence angle of $45^\circ$ inside an in-vacuum experimental chamber installed between the TDS and the high-energy spectrometer dipole, depicted in Fig.~\ref{fig:aresschematic}. The coating on the mirror serves as a dielectric boundary across which the dielectric constant changes abruptly, producing transition radiation when the beam crosses the interface. Since the mirror was positioned at $45^\circ$, the backwards CTR is emitted at $90^\circ$ to the direction of beam propagation, towards an optical viewport and was transported to the detectors outside of vacuum. This geometry allowed efficient extraction of the optical radiation while preventing direct exposure of the detectors to the electron beam. 

\begin{figure}[ht]
\centering
\includegraphics[width=0.5\textwidth]{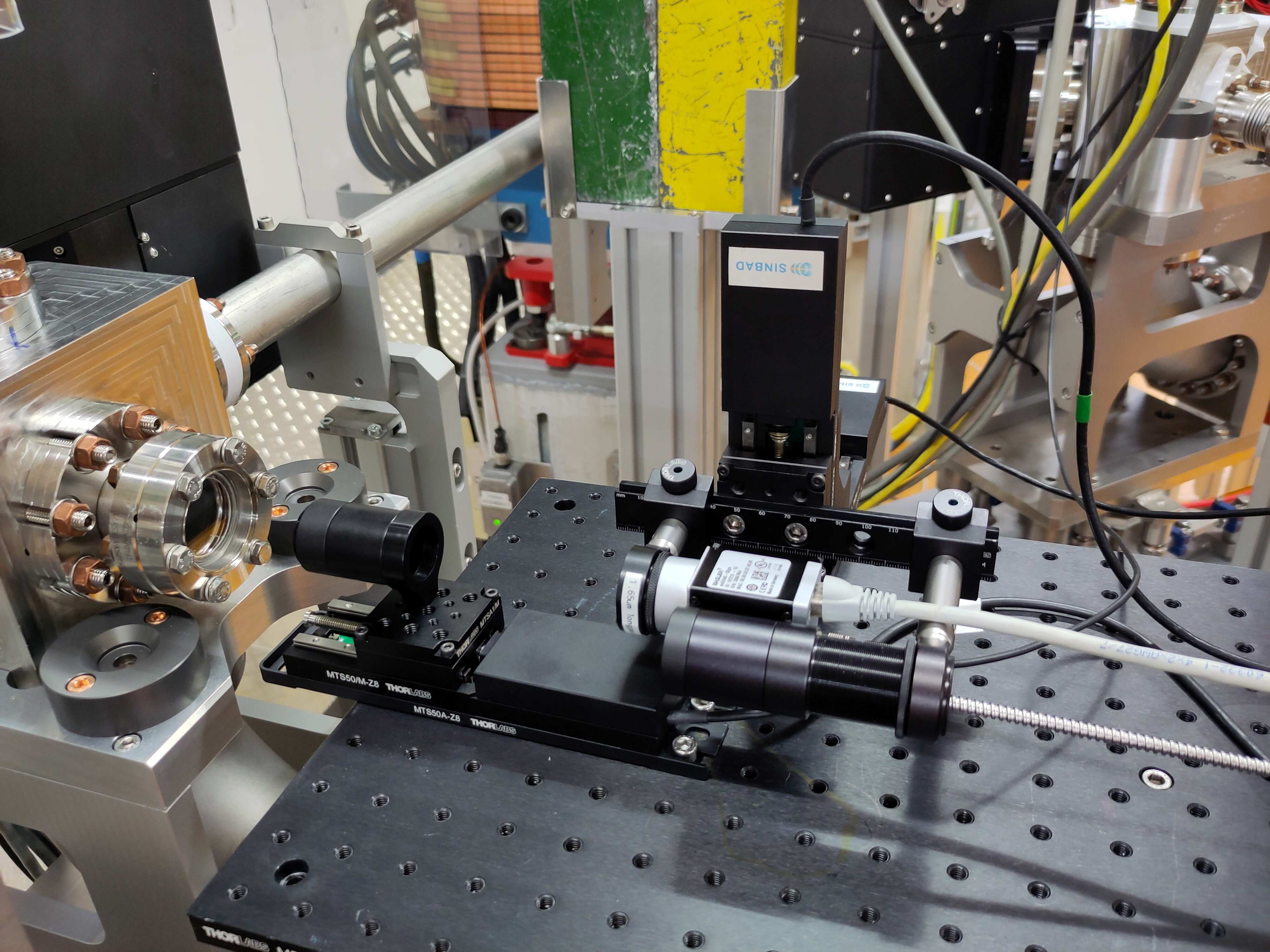}
\caption{Detector setup for COTR measurements. CMOS camera and fiber are on x-y translation stages with respect to the viewport. The lens is also on a translation stage in z with respect to the viewport.}
\label{fig:setup}
\end{figure}

\begin{figure}[ht]
\centering
\includegraphics[width=0.5\textwidth]{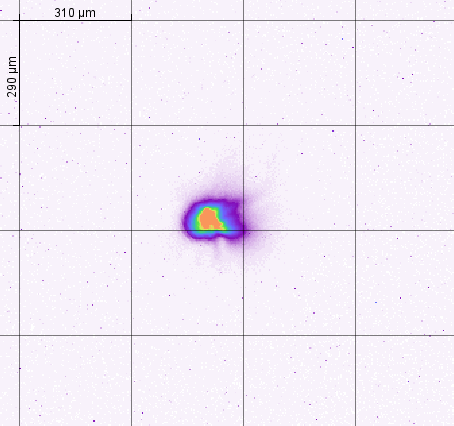}
\caption{Spatial emission of CTR as imaged by CMOS camera.}
\label{fig:spatialCTR}
\end{figure}

The spatial distribution of the radiation was recorded using a CMOS camera and is shown in Fig.~\ref{fig:spatialCTR}. Spectral measurements were performed using a fiber-coupled spectrometer covering a wavelength range of 200–1100 nm. The signal collection was optimized to the 900 um fiber entrance using a converging lens on a translation stage, to ensure optimal focusing.

For each measurement, the recorded signal was averaged over 100 consecutive electron bunches to improve the signal-to-noise ratio and reduce shot-to-shot fluctuations. The resulting spectra represents the mean optical emission from the ensemble of shots. All measurements were performed with the beamline operating under stable accelerator conditions, with the bunch compression settings fixed for each dataset.

\subsection{Theoretical model}
% ==========================================================

\label{sec:ctr_model}

To interpret the measured spectra, we model the emission as coherent transition radiation (CTR) generated when a relativistic electron bunch crosses a dielectric boundary. The calculation follows the standard decomposition into single-electron emission and bunch-enhanced coherence \cite{LihnEtAl1996,ShibataEtAl1999}. For a relativistic electron, the angular distribution of transition radiation at a planar boundary is described by the relativistic kernel
\begin{equation}
W_1(\theta)
=
A\,
\frac{\beta^{2}\sin^{2}\theta}
     {\left(1-\beta^{2}\cos^{2}\theta\right)^{2}},
\label{eq:W1_kernel}
\end{equation}
where $\theta$ is the observation angle relative to the beam axis and $A$ is a normalization constant. In comparison to experiment, this prefactor is absorbed into the overall detection efficiency, such that only the spectral shape is used to assess superradiant emission.
\\
Following the work of \cite{TerMikaelian1972,GoverEtAl2019} on CTR,
the spectral--angular emission from a bunch of $N$ electrons can be written as
\begin{equation}
\frac{d^2W}{d\omega\, d\Omega}
=
\frac{d^2W_1}{d\omega\, d\Omega}
\left[
N + N^2 |F(\omega,\theta)|^2
\right],
\label{eq:CTR_scaling}
\end{equation}
where $d^2W_1/d\omega d\Omega$ is the single-electron spectrum and 
$F(\omega,\theta)$ is the longitudinal form factor evaluated at the
appropriate phase conditions for transition radiation.

The bunch profile is modeled as Gaussian,
\begin{equation}
\rho(z)
=
\frac{1}{\sqrt{2\pi}\sigma_z}
\exp\!\left(-\frac{z^2}{2\sigma_z^2}\right),
\qquad
\sigma_z = c\sigma_t,
\qquad
\sigma_t = \frac{\tau_{\mathrm{FWHM}}}{2\sqrt{2\ln 2}},
\label{eq:gaussian}
\end{equation}
with the form factor evaluated numerically via fast Fourier transform and normalized such that $F(0)=1$.

For a bunch containing $N_e=Q/e$ electrons, the spectral–angular emission separates into incoherent and coherent contributions,
\begin{equation}
W_N(\omega,\theta)
=
W_1(\theta)
\left[
N_e
+
N_e(N_e-1)\,F^{2}(\omega,\theta)
\right].
\label{eq:WN}
\end{equation}
The first term scales linearly with charge and represents incoherent OTR. The second term scales approximately as $N_e^2$ and represents superradiant emission governed by a longitudinal form factor. 
\\
To compare with measured spectra, we integrate over the collection solid angle assuming azimuthal symmetry,
\begin{equation}
W_{\mathrm{1D}}(\omega)
=
2\pi
\int
W_N(\omega,\theta)
\sin\theta\,d\theta.
\label{eq:W1D}
\end{equation}
This yields a one-dimensional spectral prediction as a function of wavelength via the substitution $\omega=2\pi c/\lambda$.

In our experiment, transition radiation is generated when the beam strikes a $45^{\circ}$ silver mirror in vacuum. The predicted spectrum is obtained by convolving $W_{\mathrm{1D}}(\lambda)$ with the wavelength-dependent optical response of the collection system, including mirror reflectivity, lens reflectance, and spectrometer efficiency. The resulting prediction is compared directly to the measured spectra after background subtraction and normalization. The inferred bunch duration should be interpreted within the assumptions of the Gaussian longitudinal profile and finite optical bandwidth. 

% ==========================================================
\section{Results}
% ==========================================================

\begin{figure}[htbp]
\centering

% -----------------------
% Top row: (a) and (b)
% -----------------------
\begin{subfigure}[t]{0.49\textwidth}
    \centering
    \includegraphics[width=\linewidth]{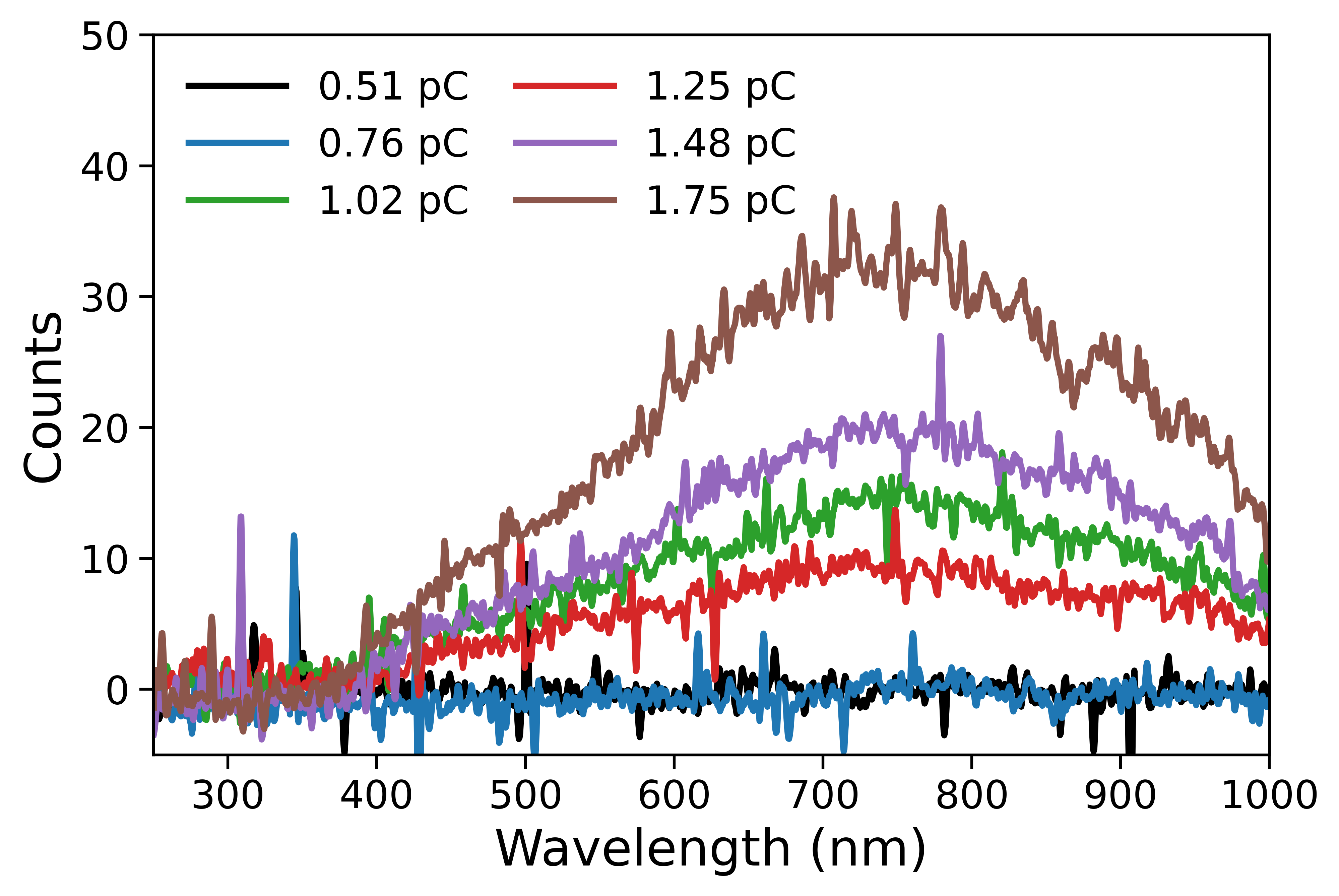}
    \caption{}
    \label{fig:spectra_scaling_a}
\end{subfigure}
\hfill
\begin{subfigure}[t]{0.49\textwidth}
    \centering
    \includegraphics[width=\linewidth]{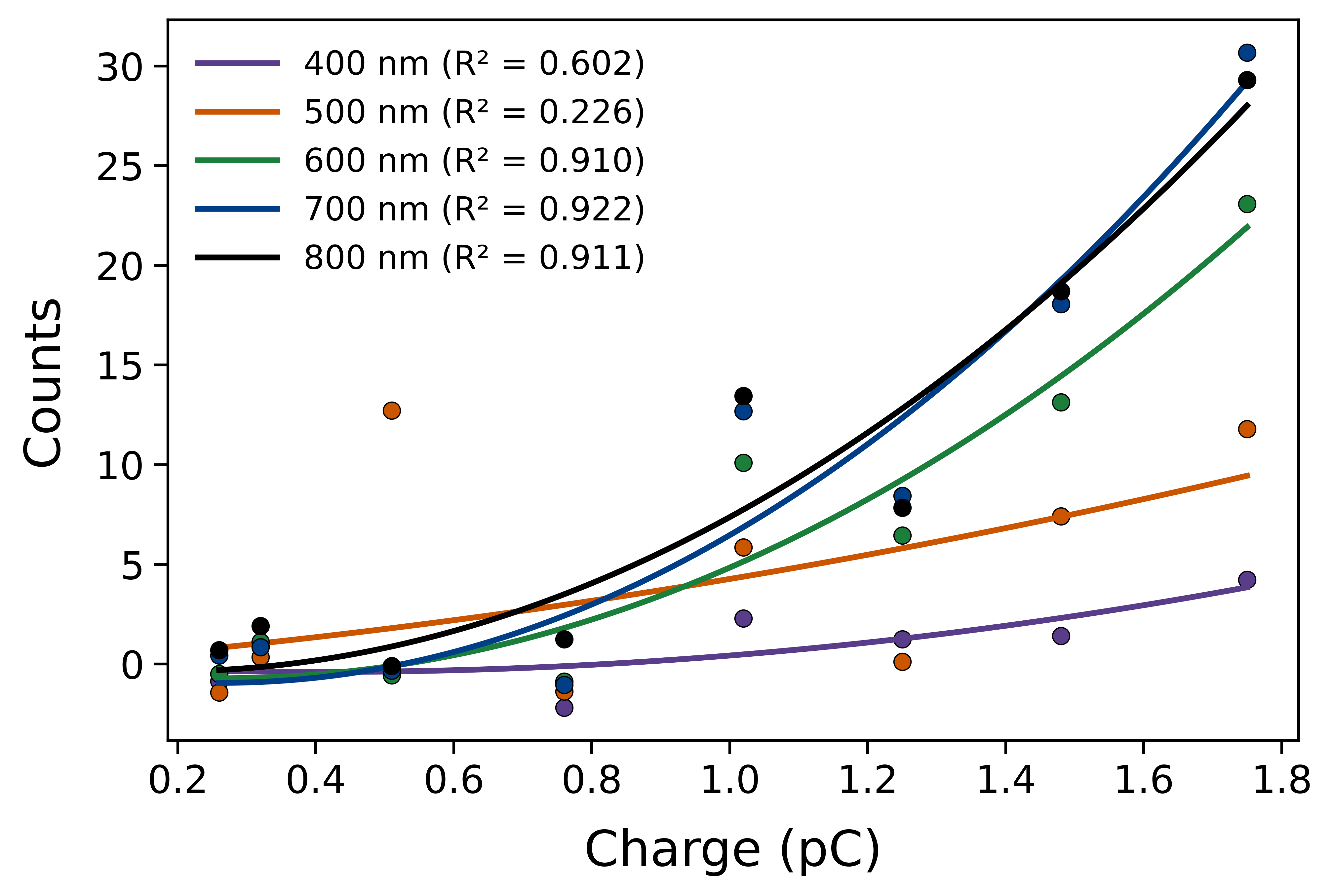}
    \caption{}
    \label{fig:spectra_scaling_b}
\end{subfigure}

\vspace{0.5em}

% -----------------------
% Bottom row: (c)
% -----------------------
\begin{subfigure}[t]{0.49\textwidth}
    \centering
    \includegraphics[width=\linewidth]{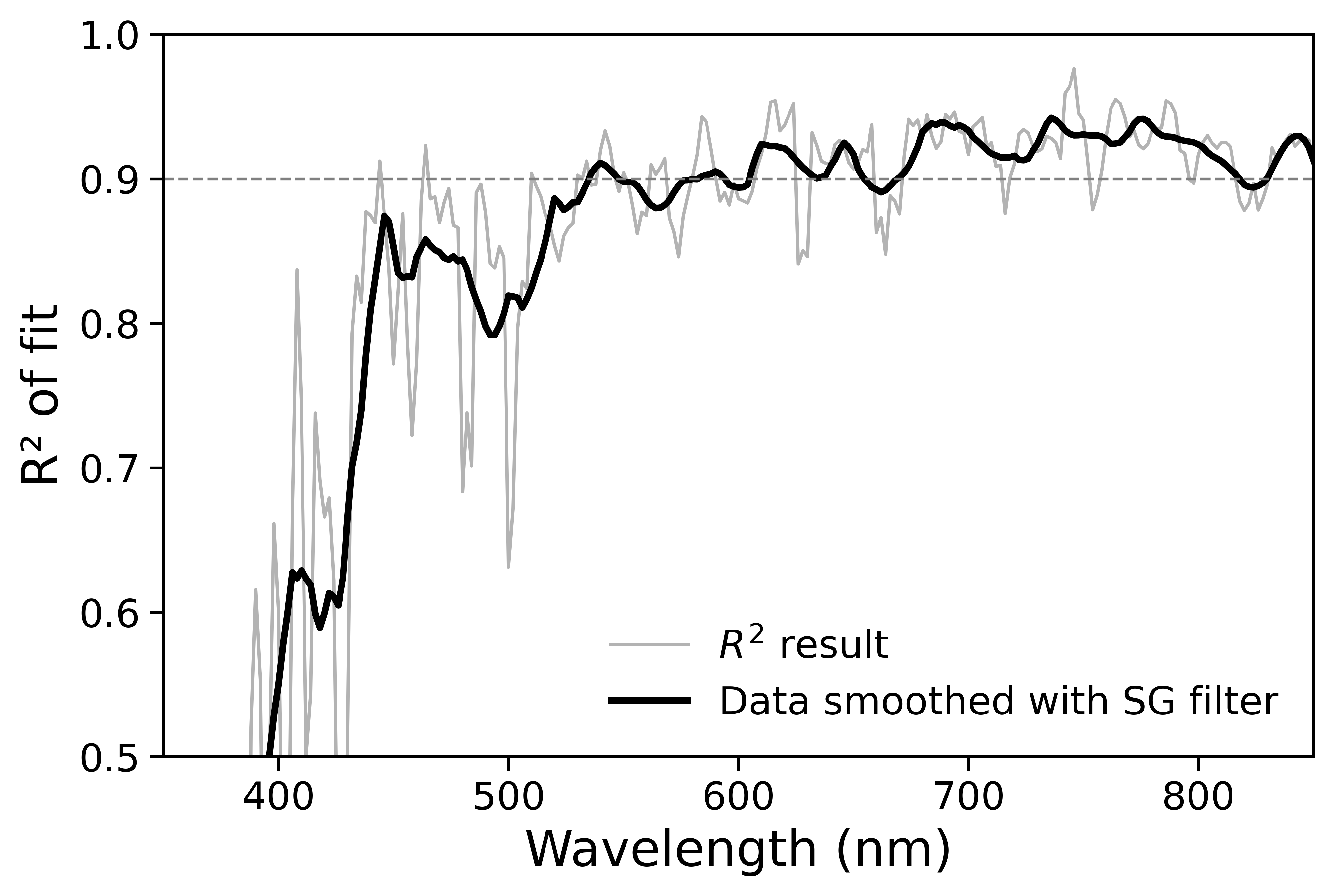}
    \caption{}
    \label{fig:spectra_scaling_c}
\end{subfigure}

\caption{Optical spectra and charge-scaling results. (a) Collected optical spectra for several bunch charges at strong compression. (b) Detector counts versus bunch charge at selected wavelengths, showing the increase in the quadratic scaling expected for coherent emission. (c) The goodness-of-fit ($R^{2}$) for the charge-scaling model $I(Q)=aQ^{2}+bQ$. The black curve shows the spectrum after smoothing with a Savitzky--Golay (SG) filter to suppress high frequency noise while preserving the spectral envelope. The dashed line indicates $R^{2}=0.9$, indicating the wavelength range where the charge scaling is well described by the quadratic model.}

\label{fig:spectra_scaling}

\end{figure}

We report the first direct observation of coherent transition radiation (CTR) extending into the visible spectral range. Access to this regime is enabled by strong longitudinal compression at the ARES beamline, producing electron bunches sufficiently short to satisfy the coherence condition at optical frequencies. Figure~\ref{fig:spectra_scaling_a} shows the measured photon spectra for several bunch charges under the strongest compression setting (TWS2 phase $=34^\circ$ with respect to the maximum momentum gain phase). The spectral intensity increases nonlinearly with charge as shown in Figure~\ref{fig:spectra_scaling_b}, consistent with coherent photon contribution. 

To quantify the scaling behavior, the photon yield at each wavelength was analyzed as a function of bunch charge in Figure~\ref{fig:spectra_scaling_b}. The charge dependence was modeled as
\begin{equation}
N_{\gamma}(\lambda,Q) = a(\lambda)\,Q^2 + b(\lambda)\,Q,
\label{eq:fit}
\end{equation}
where the quadratic term represents the coherent (superradiant) contribution and the linear term captures residual incoherent emission. The fits were constrained to pass through the origin, consistent with the physical requirement $N_{\gamma}(Q\rightarrow 0)=0$ after background subtraction.

Fig.~\ref{fig:spectra_scaling_b} shows charge scaling curves from 400--800~nm. In the 550--700~nm range, the data follow a clear quadratic dependence on bunch charge, which demonstrates superradiant emission at optical frequencies. The extracted quadratic coefficients increase across the visible range, consistent with enhanced coherence at longer wavelengths. The ratio $|b/a|$ provides a characteristic charge scale $Q_{\mathrm{cross}} \sim |b/a|$, and for 510--700~nm, $|b/a| \approx 0.5~\mathrm{pC}$, indicating that for $Q \gtrsim 0.5~\mathrm{pC}$ the emission is dominated by the coherent term. At shorter wavelengths below 500~nm, the data deviate from a quadratic relationship, consistent with reduced coherence as the wavelength approaches the longitudinal form factor cutoff. 

The observed intensity depends on both bunch charge and bunch duration, so variations in beam parameters between accelerator settings may affect the measured scaling behavior. Changes in focusing conditions can also modify the transverse form factor of the beam, influencing the intensity scaling as discussed in~\cite{KubeJINST2018}. The small deviations from monotonic charge scaling observed in the spectra likely arise from changes in beam focusing or transverse coherence.

The goodness-of-fit is summarized in Fig.~\ref{fig:spectra_scaling_c}, which shows the emergence of a coherence-dominated regime under these operating conditions. We identify this regime using the quadratic charge-scaling model, defining coherence as wavelengths where the fit satisfies $R^2 \gtrsim 0.9$. This condition is met above approximately $550~\mathrm{nm}$, indicating that in this range the emission is well described by the coherent (superradiant) contribution. For bunch charges $Q \gtrsim 1~\mathrm{pC}$, the radiation is therefore dominated by the coherent term.

%% horizontal row
%\begin{figure}[ht]
%\centering
%\includegraphics[width=\textwidth]{Figures/CTR_3panel_34deg.png}
%\caption{Spectra and charge-scaling analysis. (a) Collected optical spectra for several bunch charges at strong
%compression. (b) Photon yield versus bunch charge at selected wavelengths. (c) Wavelength-resolved goodness-of-fit
%showing the emergence of a regime consistent with quadratic scaling above $\sim 510$~nm for this operating point.}
%\label{fig:spectra_scaling}
%\end{figure}

%%%% three-panel format: top 2, bottom 1

%% Fitting the CTR spectra
\begin{figure}[ht]
\centering
\includegraphics[width=\textwidth]{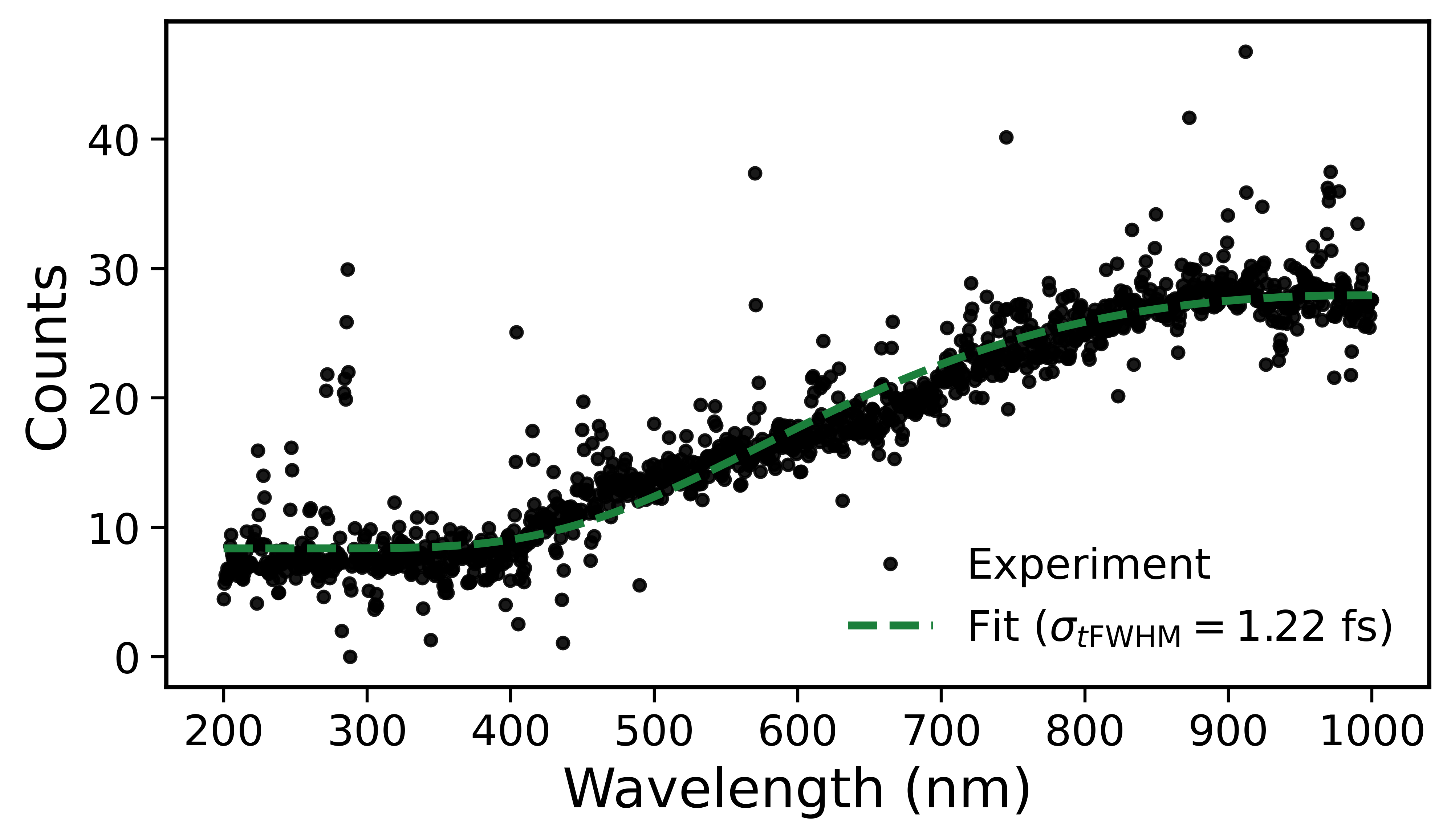}
\caption{Fit (dashed green) of the experimental CTR spectra (black) using CTR model.}
\label{fig:spectrum_fit}
\end{figure}

To estimate the bunch duration from the observed optical-frequency coherence, we assume a Gaussian longitudinal current profile. The corresponding temporal form factor satisfies
\begin{equation}
|F(\omega)|^2 = \exp\!\left[-(\omega \sigma_t)^2\right],
\end{equation}
where $\sigma_t$ is the rms bunch duration. A simple coherence criterion $\omega\sigma_t \sim 1$ implies $\sigma_t \sim 2.7\times10^{-16}~\mathrm{s}$ ($\approx 0.27~\mathrm{fs}$ rms), i.e. $\tau_{\mathrm{FWHM}} \approx 0.64~\mathrm{fs}$ for a Gaussian distribution.

We apply our CTR model described in the Methods section and fit it to data collected at a TWS2 phase of $34^\circ$ with a bunch charge of 1.48~pC over the wavelength range 200--1000~nm. The only free parameter of the fit is the length of the electron bunch. Fitting the measured CTR spectrum with the complete model yields a best fit bunch duration of $\tau_{\mathrm{FWHM}} = 1.22~\mathrm{fs}$, corresponding to $\sigma_t = \tau_{\mathrm{FWHM}}/2.355 \approx 0.52~\mathrm{fs}$, as shown in Fig.~\ref{fig:spectra_scaling}. The difference between coherence estimate and the value obtained from the spectral fit is expected. The experimentally inferred coherence is determined from quadratic charge scaling over a finite charge range and is weighted by the spectrometer response and angular collection. In contrast, the Gaussian form factor is a continuous function of $\omega\sigma_t$ and depends on the fully integrated CTR kernel. The two criteria therefore do not need to coincide numerically; however, both approaches place the emission in a regime near the 1~fs scale. Importantly, no undulator field or externally seeded density modulation is present in the beamline. The observed coherent optical emission arises from extreme longitudinal compression through well-tuned traveling wave structures followed by interaction with a dielectric boundary, providing a direct mechanism for producing optical-frequency superradiance.

Although the Gaussian model provides a useful first order reconstruction of the bunch duration, the spectral fit cannot uniquely determine fine scale structure below the inverse optical bandwidth. This substructure may arise from beam instabilities during compression. Deviations observed at the shortest wavelengths likely reflect reduced signal-to-noise ratio and possible departure from a purely Gaussian longitudinal distribution.

% ==========================================================
\section{Conclusion}
% ==========================================================

We have demonstrated coherent transition radiation extending into the visible spectrum produced by ultrashort, high-brightness electron bunches. Charge-scaling measurements reveal quadratic intensity dependence consistent with optical-frequency coherence, and the measured spectral envelope is well reproduced by a CTR model, yielding a characteristic bunch duration of $\tau_{\mathrm{FWHM}} = 1.22~\mathrm{fs}$.

These results extend transition radiation from the terahertz into the optical regime without the use of undulators or externally seeded microbunching. The observed emission arises directly from extreme longitudinal compression and boundary interaction, establishing a mechanism for generating optical-frequency superradiance in compact accelerator systems.

Beyond diagnostics, this regime provides a pathway to broadband, beam-driven coherent light sources in the visible and near-infrared, with potential applications in ultrafast spectroscopy, interferometric probing, and phase-sensitive optical measurements.

\section*{Acknowledgements}

This work was supported by the U.S. Department of Energy under Award No. DE-SC0026272, the Air Force Office of Scientific Research under Award No. FA9550-23-1-0409, and the National Science Foundation under Award No. 2431903.

% ==========================================================
% Bibliography
% ==========================================================
\bibliography{bib}

@article{GinzburgFrank1945,
  author  = {Ginzburg, V. L. and Frank, I. M.},
  title   = {Radiation of a uniformly moving electron due to its transition from one medium into another},
  journal = {J. Phys. (USSR)},
  volume  = {9},
  pages   = {353--362},
  year    = {1945}
}

@book{TerMikaelian1972,
  author    = {Ter-Mikaelian, M. L.},
  title     = {High-Energy Electromagnetic Processes in Condensed Media},
  publisher = {Wiley-Interscience},
  year      = {1972}
}

@inproceedings{FioritoRule1994,
  author    = {Fiorito, R. and Rule, D.},
  title     = {Optical transition radiation beam diagnostics},
  booktitle = {AIP Conference Proceedings},
  volume    = {319},
  pages     = {21--28},
  year      = {1994}
}

@article{GoldsmithWalsh1988,
  author  = {Goldsmith, P. and Walsh, M.},
  title   = {Coherent radiation from short electron bunches},
  journal = {Phys. Rev. Lett.},
  volume  = {60},
  pages   = {215--218},
  year    = {1988}
}

@article{Lihn1996,
  author  = {Lihn, H. C. and Wiedemann, H. and Bocek, D.},
  title   = {Time structure measurements of picosecond electron bunches using coherent transition radiation},
  journal = {Phys. Rev. E},
  volume  = {53},
  pages   = {6413--6416},
  year    = {1996},
  doi     = {10.1103/PhysRevE.53.6413}
}

@article{Shibata1999,
  author  = {Shibata, A. and others},
  title   = {Coherent transition radiation measurements of electron bunch length},
  journal = {Phys. Rev. ST Accel. Beams},
  volume  = {2},
  pages   = {062801},
  year    = {1999},
  doi     = {10.1103/PhysRevSTAB.2.062801}
}

@article{CarrEtAl2002,
  author  = {Carr, G. L. and others},
  title   = {High-power terahertz radiation from relativistic electrons},
  journal = {Nature},
  volume  = {420},
  pages   = {153--156},
  year    = {2002}
}

@article{CasalbuoniEtAl2009,
  author  = {Casalbuoni, S. and others},
  title   = {Ultrashort electron bunch length measurements using coherent transition radiation},
  journal = {Phys. Rev. ST Accel. Beams},
  volume  = {12},
  pages   = {030705},
  year    = {2009},
  doi     = {10.1103/PhysRevSTAB.12.030705}
}

@article{LumpkinEtAl2006,
  author  = {Lumpkin, A. H. and others},
  title   = {Overview of coherent transition radiation diagnostics},
  journal = {Nucl. Instrum. Methods Phys. Res., Sect. A},
  volume  = {557},
  pages   = {59--66},
  year    = {2006}
}

@article{HuangEtAl2004,
  author  = {Huang, Z. and others},
  title   = {Observation of microbunching instability in a linac-driven free-electron laser},
  journal = {Phys. Rev. ST Accel. Beams},
  volume  = {7},
  pages   = {074401},
  year    = {2004},
  doi     = {10.1103/PhysRevSTAB.7.074401}
}

@article{Ferrario2010,
  author  = {Ferrario, M. and others},
  title   = {Experimental demonstration of emittance compensation with velocity bunching},
  journal = {Phys. Rev. Lett.},
  volume  = {104},
  pages   = {054801},
  year    = {2010},
  doi     = {10.1103/PhysRevLett.104.054801}
}

@article{Xiang2012,
  author  = {Xiang, D. and others},
  title   = {Generation of ultrashort electron bunches using linear and nonlinear compression},
  journal = {Phys. Rev. ST Accel. Beams},
  volume  = {15},
  pages   = {050707},
  year    = {2012},
  doi     = {10.1103/PhysRevSTAB.15.050707}
}

@article{Musumeci2010,
  author  = {Musumeci, P. and Moody, J. T. and Scoby, C. M. and Gutierrez, M. S. and Rosenzweig, J. B.},
  title   = {High-quality single-shot ultrafast electron diffraction from a compact photoinjector},
  journal = {Phys. Rev. Lett.},
  volume  = {104},
  pages   = {084801},
  year    = {2010},
  doi     = {10.1103/PhysRevLett.104.084801}
}

@article{Rosenzweig2018,
  author  = {Rosenzweig, J. B. and others},
  title   = {Ultrahigh brightness electron beams from cryogenic RF photoinjectors},
  journal = {Nucl. Instrum. Methods Phys. Res., Sect. A},
  volume  = {909},
  pages   = {463--469},
  year    = {2018}
}

@article{Papadopoulos2020,
  author  = {Papadopoulos, D. A. and others},
  title   = {Cryogenic radio-frequency photoinjectors for ultrahigh brightness beams},
  journal = {Phys. Rev. Accel. Beams},
  volume  = {23},
  pages   = {063401},
  year    = {2020},
  doi     = {10.1103/PhysRevAccelBeams.23.063401}
}

@article{Tantawi2020,
  author  = {Tantawi, S. G. and others},
  title   = {Breakdown-resistant high-gradient cryogenic copper accelerating structures},
  journal = {Phys. Rev. Accel. Beams},
  volume  = {23},
  pages   = {092001},
  year    = {2020},
  doi     = {10.1103/PhysRevAccelBeams.23.092001}
}

@article{BrownellWalshDoucas1998,
  author  = {Brownell, J. H. and Walsh, J. and Doucas, G.},
  title   = {Spontaneous Smith--Purcell radiation described through induced surface currents},
  journal = {Phys. Rev. E},
  volume  = {57},
  pages   = {1075--1083},
  year    = {1998},
  doi     = {10.1103/PhysRevE.57.1075}
}

@article{GoverEtAl2019,
  author  = {Gover, A. and others},
  title   = {Theory of superradiant emission from bunched electron beams},
  journal = {Rev. Mod. Phys.},
  volume  = {91},
  pages   = {035003},
  year    = {2019},
  doi     = {10.1103/RevModPhys.91.035003}
}

@inproceedings{BurkartEtAl2022,
  author    = {Burkart, F. and others},
  title     = {The ARES Linac at DESY},
  booktitle = {Proceedings of LINAC2022},
  pages     = {691--694},
  year      = {2022},
  note      = {THPOJO01},
  doi       = {10.18429/JACoW-LINAC2022-THPOJO01}
}

@article{LihnEtAl1996,
  author  = {Lihn, H. C. and Kung, P. H. and Wiedemann, H. and Bocek, D.},
  title   = {Time structure measurements of picosecond electron bunches using coherent transition radiation},
  journal = {Phys. Rev. E},
  volume  = {53},
  pages   = {6413--6416},
  year    = {1996},
  doi     = {10.1103/PhysRevE.53.6413}
}

@article{ShibataEtAl1999,
  author  = {Shibata, A. and others},
  title   = {Coherent transition radiation measurements of electron bunch length},
  journal = {Phys. Rev. ST Accel. Beams},
  volume  = {2},
  pages   = {062801},
  year    = {1999},
  doi     = {10.1103/PhysRevSTAB.2.062801}
}

@article{Lumpkin2002PRL,
  author  = {Lumpkin, A. H. and Rule, D. W. and Downer, M. C.},
  title   = {Evidence for Microbunching-Induced Coherent Optical Transition Radiation in a SASE Free-Electron Laser},
  journal = {Physical Review Letters},
  volume  = {88},
  number  = {23},
  pages   = {234801},
  year    = {2002},
  doi     = {10.1103/PhysRevLett.88.234801}
}

@article{Lumpkin2009PRSTAB,
  author  = {Lumpkin, A. H.},
  title   = {Characterization and Mitigation of Coherent Optical Transition Radiation from Compressed Electron Beams},
  journal = {Physical Review Special Topics - Accelerators and Beams},
  volume  = {12},
  pages   = {080702},
  year    = {2009},
  doi     = {10.1103/PhysRevSTAB.12.080702}
}

@article{Lumpkin2020PRL,
  author  = {Lumpkin, A. H. and et al.},
  title   = {Observation of Coherent Optical Transition Radiation Interference Patterns from Microbunched Electron Beams},
  journal = {Physical Review Letters},
  volume  = {125},
  pages   = {014801},
  year    = {2020},
  doi     = {10.1103/PhysRevLett.125.014801}
}

@phdthesis{JasterMerz2023Thesis,
    author = {Jaster-Merz, S. M.},
    title = {High-dimensional and ultra-sensitive diagnostics for electron beams},
    school = {Universit\"at Hamburg},
    year ={2023}
}

@inproceedings{KuropkaEtAl2022,
  author    = {Kuropka, W. and others},
  title     = {Commissioning of a movable bunch compressor for sub-fs electron bunches},
  booktitle = {Proceedings of LINAC2022},
  pages     = {695--698},
  year      = {2022},
  note      = {THPOJO02},
  doi       = {10.18429/JACoW-LINAC2022-THPOJO02}
}

@article{KubeJINST2018,
  author  = {Kube, G. and Potylitsyn, A.P.},
  title   = {Coherent backward transition radiation from sub-fs “pancake-like” bunches as a tool for beam diagnostics},
  journal = {J. Intrum.},
  volume  = {13},
  pages   = {CO2055},
  year    = {2018},
  doi     = {10.1088/1748-0221/13/02/C02055}
}

@misc{Grudiev2016,
  author = {Grudiev, Alexej},
  title = {Design of compact high power RF components at X-band},
  year = {2016},
  url = {https://cds.cern.ch/record/2158484},
  eprint = {CLIC-Note-1067}
}

\end{document}